\documentclass[checkin,showpacs,psfig,aps,pra]{revtex4}

\newcommand{\ba}{\begin{eqnarray}}
\newcommand{\ea}{\end{eqnarray}}

\usepackage{graphicx,color}

\newcommand{\Cr}{$\rm CrO_{2}$ }

\newcommand{\Cre}{$\rm CrO_{2}$}

\newcommand{\be}{\begin{equation}}
\newcommand{\ee}{\end{equation}}

\newcommand{\clr}{}

\newcommand{\et}{{\it et al. }}

\newcommand{\bn}{\begin{enumerate}}
\newcommand{\en}{\end{enumerate}}

\begin{document}

\title{Harmonic generation {predominantly} from a single spin channel
  in a half metal}

\author{G. P. Zhang$^*$} \affiliation{Department of Physics, Indiana State
 University, Terre Haute, Indiana 47809, USA}

\author{Y. H. Bai}
\affiliation{Office of Information Technology, Indiana State
  University, Terre Haute, Indiana 47809, USA}

\date{\today}

\begin{abstract}
{Harmonic generation in atoms and molecules has reshaped our
  understanding of ultrafast phenomena beyond the traditional
  nonlinear optics and has launched attosecond physics. Harmonics from
  solids represents a new frontier, where both majority and minority
  spin channels contribute to harmonics.}  This is true even in a
ferromagnet whose electronic states are equally available to optical
excitation.  Here, we demonstrate that harmonics can be generated
{mostly} from a single spin channel in half metallic chromium
dioxide. {An energy gap in the minority channel greatly reduces
  the harmonic generation}, so harmonics predominantly emit from the
majority channel, with a small contribution from the minority
channel. However, this is only possible when the incident photon
energy is well below the energy gap in the minority channel, so all
the transitions in the minority channel are virtual. The onset of the
photon energy is determined by the transition energy between the
dipole-allowed transition between the O-$2p$ and Cr-$3d$
states. Harmonics {mainly} from a single spin channel can be
detected, regardless of laser field strength, as far as the photon
energy is below the minority band energy gap. This prediction should
be tested experimentally.
\end{abstract}


\maketitle

\section{Introduction}

Rapid developments of high harmonic generation (HHG) in atoms and
molecules \cite{mcpherson1987,ferray1988} have opened new frontiers in
material research and ultrafast science.  Starting from 1990s,
interest has shifted from atoms, molecules to nanostructure and actual
solids
\cite{farkas1992,vonderlinde1995,prl05,ganeev2009a,ghimire2011,luu2015,garg2016,nd}.
     {\clr Solids are much more complex, and they demand more from
       HHG. For instance, can HHG directly probe band structures in a
       solid \cite{vampa2015a,vampa2015b,prb20}?}  Developing harmonics
       into a tool to explore properties in topological materials such
       as Bi$_2$Se$_3$ \cite{jia2019,baykusheva2021}, MnBi$_2$Te$_4$
       \cite{jia2020b}, MoS$_2$ \cite{jia2020a} and spin-orbit coupled
       systems \cite{lysne2020} is particularly important, given that
       vast majority of devices are solids.  Up to now, most of
     materials investigated so far are nonmagnetic and most of them
     are insulators and semiconductors
     \cite{ghimire2011,garg2016,you2017}.  In ferromagnets, both
     majority and minority spin channels contribute to harmonic
     signals \cite{nc18}, so it is difficult to disentangle their
     respective contributions cleanly \cite{takayoshi2019}.  In
     antiferromagnets, one encounters the same difficulty where two
     sublattices with opposite spin orientations both contribute to
     harmonic signals \cite{Tancogne-Dejean2018}. So far, pure spin
     polarized harmonics in solids have not materialized, but are in
     huge demand in materials where spin dynamics is important. In
     laser-induced ultrafast demagnetization \cite{eric,ourreview},
     one needs to know how differently two spin channels evolve with
     time.  THz emission was reported in Cr(30 $\rm\AA$)/Ni(42
     $\rm\AA$)/Cr(70 $\rm \AA$) films \cite{eric2004}, and more
     recently in Co/ZnO/Pt and Co/Cu/Pt multilayers \cite{li2018} and
     MgO/Fe/MgO thin films \cite{zhang2020}. THz emission only allows
     one to access magnetization on a picosecond time scale.  Being
     able to detect the spin signal from harmonics beyond THz emission
     potentially provides a new way to investigate spin wave dynamics
     \cite{zhang2002,muller2009} on a few hundred femtoseconds.

In this paper, we demonstrate harmonic generation {mostly} from a
single spin channel in half metallic \Cr \cite{schwarz1986,coey}.  \Cr
has no energy gap in the majority spin channel, but has a gap in the
minority one.  The minority state is away from the Fermi level and the
spin-orbit coupling is weak, leading to nearly 100\% spin polarization
\cite{soulen1998}.  These interesting properties manifest themselves
in harmonic generation. The energy gap in the minority band limits the
efficiency of harmonic generation, but the effect on each order of
harmonics is different. The fundamental harmonic, the first harmonic,
is weakly affected, and has no strong dependence on the incident laser
photon energy. A much stronger effect is found in higher orders of
harmonics. With the laser parameters used, we find that upon
excitation of a 0.4-eV pulse, the fifth harmonic from the majority
channel is 60 times stronger than that from the minority signal,
making it a nearly pure single-spin channel harmonic. This difference
disappears when a higher photon energy is used. Such a dramatic effect
is closely related to the dipole-allowed transitions between the
oxygen $2p$ states and chromium $3d$ state, where the strongest
transition is slightly above 2 eV. It is generally believed that
harmonic generation does not have the spin selectivity.  Our finding
here demonstrates that optical harmonics can be a powerful tool to
access spin information in half-metals, where harmonics are {\clr
  largely} emitted from a single spin channel.

{The rest of the paper is arranged as follows. In Sec. II, we present
our theoretical formalism. The results and discussions are in Sec.
III. We conclude this paper in Sec. IV. }

\section{Theoretical formalism}

\newcommand{\ik}{i{\bf k}}
\newcommand{\jk}{j{\bf k}}

\newcommand{\rr}{{\bf r}}

We employ the first-principles density functional theory and couple it
to the time-dependent Liouville equation. This formalism represents an
important alternative to the time-dependent density functional theory
and rigorously respects the Pauli exclusion principles \cite{jpcm16}.
Here in brief, we solve the the spin-polarized Kohn-Sham equation
\cite{wien2k,blaha2020,np09,prb09}, \be \left
     [-\frac{\hbar^2\nabla^2}{2m_e}+V_{ne}+V_{ee}+V_{xc} \right
     ]\psi_{\ik}^\sigma(\rr)=E_{\ik}^\sigma \psi_{\ik}^\sigma
     (\rr), \label{ks} \ee where $\sigma$ is the spin index, $m_e$ is
     the electron mass, the terms on the left-hand side represent the
     kinetic energy, nuclear-electron attraction, electron-electron
     Coulomb repulsion and exchange correlation (generalized gradient
     approximation \cite{pbe}), respectively.
     $\psi_{\ik}^\sigma(\rr)$ is the Bloch wavefunction of band $i$
     for spin $\sigma$ at crystal momentum ${\bf k}$, and
     $E_{\ik}^\sigma$ is the band energy. {Our two spin channels
       are not mixed because of weak spin-orbit coupling
       \cite{soulen1998}. This allows us to cleanly investigate the
       contributions from these two spin channels}.  We use the
     full-potential augmented plane wave method as implemented in the
     Wien2k code \cite{wien2k}, where within the Muffin-tin sphere
     atomic basis functions are used and in the interstitial region a
     plane wave is used. These two basis functions are matched both in
     value and slope at the Muffin-tin boundary.

To investigate high harmonic generation, we solve the Liouville
equation in real time. Different from prior studies \cite{prb19}, we
have to solve the equation for each spin channel $\sigma$, \be
\frac{\partial \rho_{\bf k}^\sigma}{\partial t}= \frac{1}{i\hbar}
     [H_0^\sigma+H_I^\sigma,\rho_{\bf k}^\sigma] -\frac{ \rho_{\bf
         k}^\sigma- \rho_{\bf k}^\sigma(0)}{T_{1(2)}}, \label{decay}
     \ee so the computational cost doubles.  Here $\rho_{\bf
       k}^\sigma$ is the time-dependent density matrix with spin
     $\sigma$ at the ${\bf k}$ point and $\rho_{\bf k}^\sigma(0)$ is
     its initial value. {The second term on the right side
       describes the decay of the density matrix to its initial value
       \cite{shen}.  For a diagonal density matrix element $\rho_{{\bf
           k};ii}^\sigma$, the second term becomes $-[\rho_{{\bf
             k};ii}^\sigma - \rho_{{\bf k};ii}^\sigma(0)]/T_1 $, where
       $i$ is the band index and $T_1$ is called the longitudinal
       relaxation time. For an off-diagonal density matrix element
       $\rho_{{\bf k};i\ne j}^\sigma$, the second term is $-\rho_{{\bf
           k};i\ne j}^\sigma /T_2 $, where $i$ and $j$ are the band
       indices and $T_2$ is called the transverse relaxation time.
       Note that the initial off-diagonal matrix elements are
       zero. $T_1$ and $T_2$ are different from those in NMR where
       $T_1$ is used for the $z$ component of magnetization $M_z$ and
       $T_2$ is used for the $x$ and $y$ components \cite{kittel}.}
     $H_0$ is the field-free system Hamiltonian.  $H_I^\sigma$
     represents the interaction between the laser field and system, $
     H_I^\sigma=-\sum_{{\bf k};i,j} \rho_{{\bf k};i,j}^\sigma {\bf
       P}_{{\bf k};j,i}^\sigma \cdot {\bf A}(t)$.  {${\bf
         P}_{{\bf k};ij}^\sigma$ is the momentum matrix element
       between bands $i$ and $j$ at the ${\bf k}$ point, and is
       calculated within the Wien2k code, but these matrix elements
       must be converted properly before use (see the details in
       Ref. \cite{ourbook})}.  ${\bf A}(t)$ is the vector potential, $
     {\bf A}(t)= A_0 \sin^2(t/\tau) (\cos(\omega t) \hat{x} \pm
     \sin(\omega t) \hat{y}), $ where $A_0$ is the amplitude of the
     vector potential in units of Vfs$/\rm\AA$ \cite{nc18}, $\hat{x}$
     and $\hat{y}$ are the unit vectors along the $x$ and $y$ axes,
     respectively, $t$ is the time, $\omega$ is the carrier frequency,
     + and $-$ refer to the left ($\sigma^-$) and right ($\sigma^+$)
     circularly polarized light within the $xy$ plane, respectively.
     $\tau$ is the laser pulse duration.  We choose the duration
     $\tau$ to be 64 cycles of laser period \cite{prl05}. This is
     sufficient to resolve harmonics at different orders. If the laser
     polarization is within the $yz$ plane, those unit vectors must be
     changed properly.

\section{Results and discussions}

 \Cr crystallizes in the rutile structure, with the space group $\rm
 P4_2/mnm$.  The Cr atoms form a body-centered tetragonal lattice.  If
 we orient the structure in such a way that the Cr atom is at the
 center of the oxygen distorted octahedron, then two apex oxygen atoms
 $O_1$ and $O_2$ are placed along the vertical direction. Four
 equatorial oxygen atoms $O_3$,..., and $O_6$, whose Wyckoff positions
 are $(u,u,0)$, $(u,u,1)$, $(1-u,1-u,0)$, and $(1-u,1-u,1)$, are
 placed in the horizontal plane. The distance between the body center
 Cr atom and $O_1$ and $O_2$ is shorter, $d_a=1.8962 \rm \AA$, and
 that between the body center Cr atom and $O_3$,..., and $O_6$ is
 longer, $d_e=1.9087 \rm \AA$, both of which are computed from
 $d_a=\sqrt{2}ua$ and $d_e=\sqrt{2(\frac{1}{2}-u)^2a^2+(c/2)^2}$
 \cite{sorantin1992}. There are two distances between neighboring
 equatorial O atoms, $c$ and $\sqrt{2}(2u-1)a$, and normally they
 differ, so one does not have an ideal octahedron.

 Our ${\bf k}$ mesh is $29\times 29 \times 45$, with 2760 points in the
 irreducible Brillouin zone.  {The product of planewave cutoff $K_{\rm
     max}$ and Muffin-tin radius $R_{\rm MT}$, $R_{\rm MT}K_{\rm
     max}$, determines the accuracy of our calculation.  Our $R_{\rm
     MT}K_{\rm max}$ is 9, which is more than enough for our
   purpose. $R_{\rm MT}(Cr)$ is 1.89 a.u. and $R_{\rm MT}(O)$ is 1.67
   a.u.  We optimize the internal coordinates of atoms, and we compare
   our results with the experimental ones in Table \ref{tab}.} We see
 that the agreement with the experiment is excellent.  Our theoretical
 spin moment for formula unit is 2.0 $\mu_B$, which agrees with the
 experimental result of 2.0 $\mu_B$ nearly perfectly and prior
 theoretical calculations \cite{schwarz1986,sorantin1992}.

\begin{figure}

\includegraphics[angle=0,width=1\columnwidth]{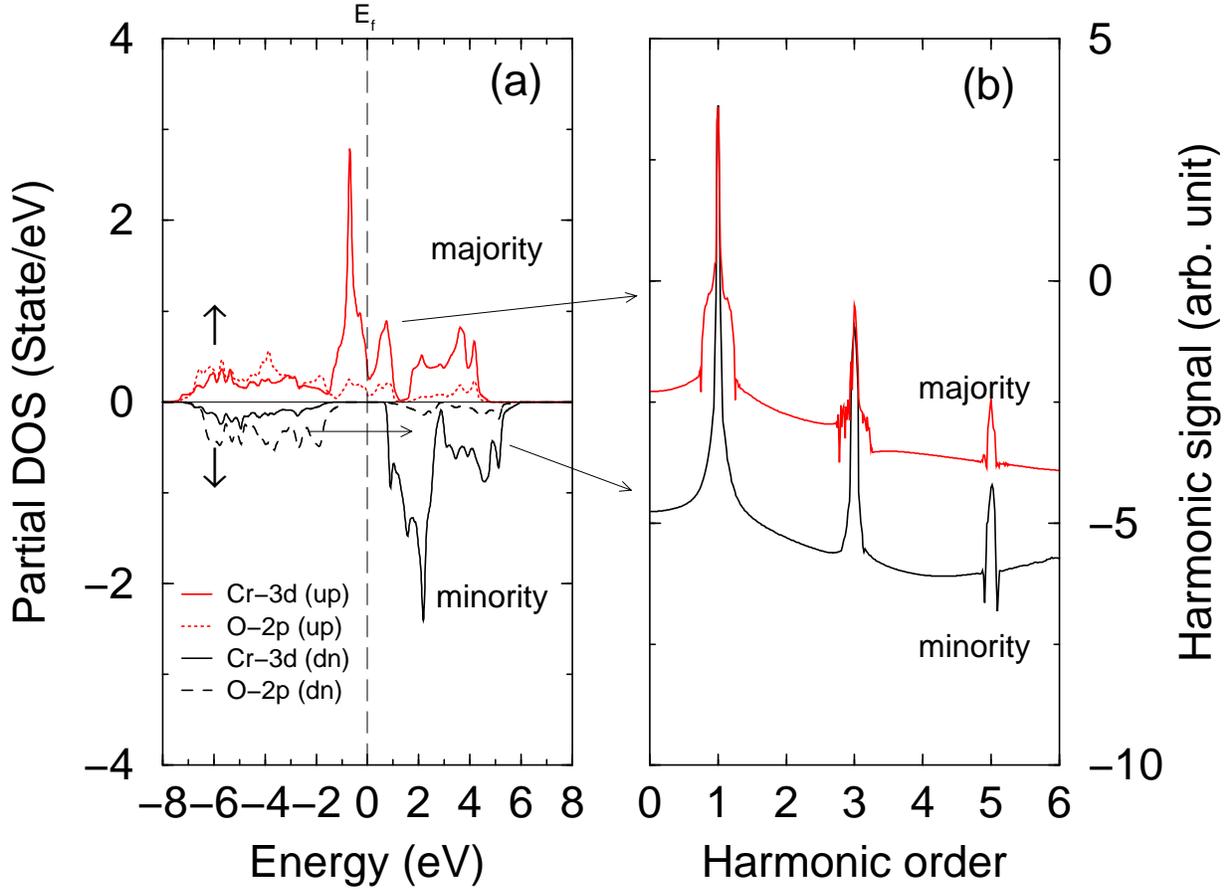}
\caption{ (a) The partial density of states (PDOS) for Cr-$3d$ and
  O-$2p$ states for the majority (thick solid and dotted lines on the
  positive axis) and minority spins (solid and dashed lines on the
  negative axis). The vertical dashed line denotes the Fermi energy.
  (b) {Logarithmic scale of harmonics from the majority and
    minority spin channels.} Here we use the laser duration of 64
  cycles of laser period. The photon energy is 0.4 eV and the laser
  field amplitude $A$ is 0.03 $\rm Vfs/\AA$. A circularly polarized
  pulse is used.  }
\label{fig1}
\label{pdos}
\end{figure}

Figure \ref{pdos}(a) displays the partial density of states (PDOS) for
the majority (positive value) and minority (negative value) spins.
The solid line and dotted line denote the Cr-$3d$ and O-$2p$ majority
states, respectively. The vertical dashed line represents the Fermi
energy which is set at 0. One sees that both Cr-$3d$ and O-$2p$ states
cross the Fermi level, which is why \Cr is a metal. Each cell has 28
valence majority electrons and 24 minority electrons.  The Fermi
energy nearly cuts at the local minimum of Cr-$3d$ states, stabilizing
\Cr \cite{sorantin1992}.  The minority PDOS is quite different (see
the solid and dashed lines on the negative axis). At the Fermi level,
there is no DOS, with a band gap of 1.43 eV \cite{schwarz1986}. These
features are fully consistent with prior studies
\cite{schwarz1986,sorantin1992,mazin1999}.  What is important for
harmonic generation is that the valence band is dominated by the
O-$2p$ state, while the conduction band is dominated by the Cr-$3d$
state. This constitutes an ideal case for dipole-allowed transitions.

We find there is no major difference between left and right circularly
polarized light for our system, so in the following we use $\sigma^+$,
with the photon energy 0.4 eV and field amplitude $A_0=0.03$ Vfs/$\rm
\AA$. 
%, i. e. {$1.162 \times 10^9\rm W/cm^2$}
Our interested observable is the expectation value of the momentum
operator, $\langle {\bf P}^\sigma(t) \rangle =\langle \sum_{{\bf
    k};i,j} \rho_{{\bf k};i,j}^\sigma {\bf P}_{{\bf
    k};j,i}^\sigma\rangle, $ where the summation is over the Brillouin
zone and band states. In our calculation, we include all the states
from Cr-$3s$ states (energy at $-4.89$ Ry for spin up and $-4.73$ Ry
for spin down), Cr-$3p$ state (energy at $-2.79$ Ry for spin up and
$-2.64$ Ry for spin down), O-$2s$ (-0.95 Ry), all the way up to 4.95
Ry, with 10 Ry energy window width covering all the possible channels
for excitation. To compute harmonic spectra, we Fourier transform
$\langle {\bf P}^\sigma(t) \rangle$ into the frequency domain
\cite{nc18,prl05,jia2020a}, \be {\bf P}(\Omega)=\int_{-\infty}^{\infty} {\bf
  P}(t) {\rm e}^{i\Omega t} {\cal W}(t) dt, \ee where ${\cal W}(t)$ is
the window function \cite{nc18}.

 Figure \ref{fig1}(b) shows harmonic signals from the majority (top)
 and minority (bottom) spin channels, respectively.  We see that all
 the harmonics appear at odd orders because our system has an
 inversion symmetry. The majority spin channel has a stronger harmonic
 signal than the minority spin channel, and it has a broader peak.
 This difference is expected from the PDOS. Recall that the minority
 band has an energy gap, so with the photon energy of 0.4 eV, the
 excitation is virtual.  The majority band has no such gap, so
 electrons in the valence band can successively excite into conduction
 bands and then emit harmonics. 

We notice a crucial difference when we compare the 3rd harmonic signal
with the 5th harmonic signal from the majority and minority spin
channels. The difference in the 5th harmonic between the majority and
minority spin channels is much larger.  A very weak 5th harmonic in
the minority channel is interesting because it potentially provides an
opportunity to generate harmonics from a single spin channel. We note
that in a nonmagnetic material \cite{jia2019}, there is no difference
in harmonics between spin up and spin down channels. In ferromagnets,
this difference is also small because each channel is able to generate
harmonics with a small difference in strength
\cite{nc18,jia2020a,jia2020b}.  A challenge in ferromagnetic metals is
that the electron deexcitation is extremely fast \cite{nc18}, so it is
difficult to detect those harmonics. CrO$_2$ is very unique. It is
known that laser-induced ultrafast demagnetization lasts very long
\cite{zhang2006}, partly because its spin-orbit coupling is very
weak. This prolonged process is advantageous to detect harmonic
signals.

\begin{figure}
\includegraphics[angle=0,width=0.8\columnwidth]{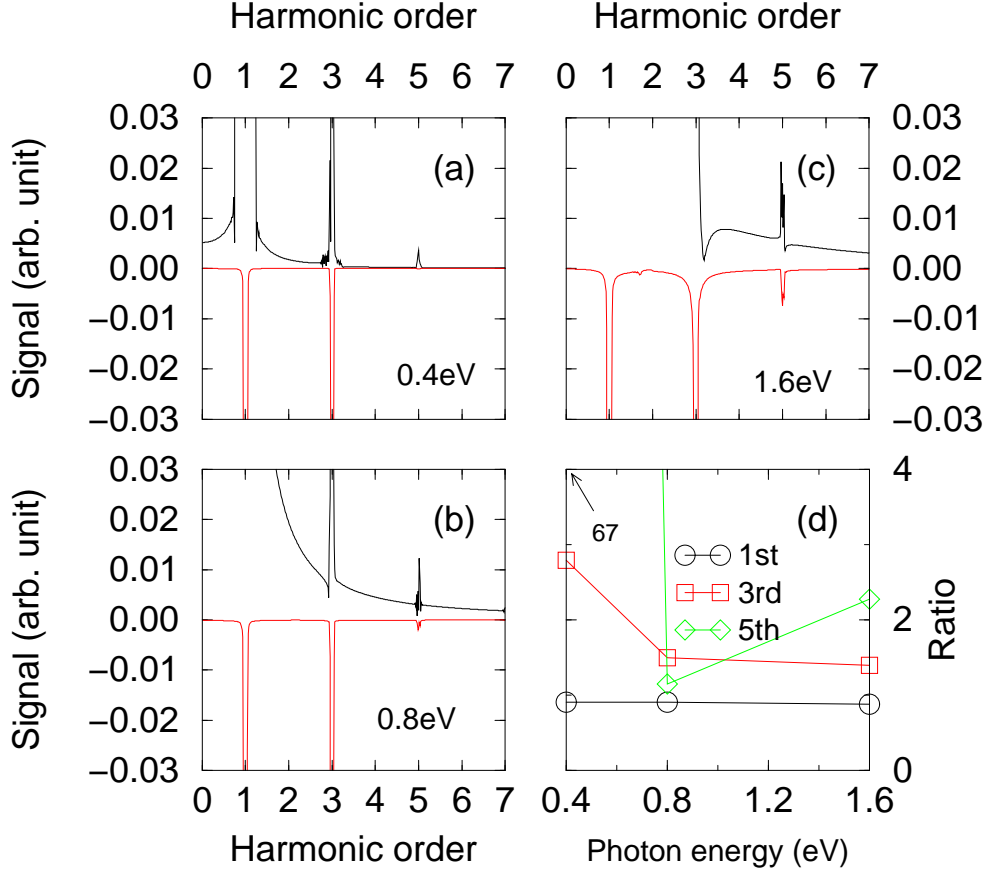}
\caption{ Dependence of harmonics on photon energies $\hbar\omega$ (a)
  0.4 eV, (b) 0.8 eV and (c) 1.6 eV. We choose the vector potential
  $A_0=0.03 \rm Vfs/\AA$ and the duration of the laser pulse is 64
  cycles of laser period. Harmonics from the minority channel are
  plotted on the negative axis.  (a) At 0.4 eV, the majority and
  minority spin channels produce a similar signal strength at the 1st
  and 3rd orders, but differ a lot at the 5th order, where the
  majority dominates over the minority spin channel. This presents an
  opportunity to generate harmonic mainly from a single spin
  channel. As we increase photon energy to 0.8 eV (b), the minority
  5th harmonic increases its signal strength, so it starts to compete
  with that from the majority channel. This trend continues as we
  increase the photon energy to 1.6 eV (c), where the majority and
  minority channels have a similar strength.  (d) Ratio of the
  majority harmonic signal to the minority harmonic signal as a
  function of photon energy for the first (circles), third (squares)
  and fifth (diamonds) harmonics. At 0.4 eV, the ratio for the fifth
  harmonic reaches 67.  }
\label{fig2}
\end{figure}

Next we investigate how laser parameters affect harmonic generation in
CrO$_2$. We increase the photon energy to 0.8 eV, while keeping the
rest of laser parameters unchanged. Figure \ref{fig2}(b) shows that
the 5th harmonic in the minority channel at 0.8 eV is slightly smaller
than that in the majority channel, very different from that at 0.4 eV
(see Fig. \ref{fig2}(a)). Energetically, when the photon energy
$\hbar\omega$ is 0.4 eV, $5\hbar\omega=2$ eV is slightly above the gap
energy of 1.43 eV. Because a transition from the O-$2p$ to Cr-$3d$
does not occur immediately at this band gap energy, one is still in
the virtual excitation regime. This situation is important because not
all half metals have this property. For instance, in another
half-metal, Mn$_2$RuGa does not have this unique feature, so its
harmonic generation from the majority channel is similar to that in
the minority channel. With $\hbar\omega= 0.8$ eV, $5\hbar\omega=4$ eV
is well into the real excitation regime, so the band gap suppression
is no longer at work. This is further verified when we increase the
incident photon energy to 1.6 eV. Figure \ref{fig2}(c) shows that
harmonic strength in the minority and majority channels is almost
exactly the same, so harmonics are generated from both spin
channels. Figure \ref{fig2}(d) compares the harmonic signal ratio of
the majority harmonic signal to the minority harmonic signal as a
function of incident photon energy. We see that in general higher
order harmonics have a much stronger variation. The ratio is larger
when the incident photon energy is small. At 0.4 eV, the fifth
harmonic has a ratio of 67 with $A_0=\rm 0.03Vfs/\AA$.  This photon
energy window predicted here is critical to future experiments.

\begin{figure}
\includegraphics[angle=0,width=0.8\columnwidth]{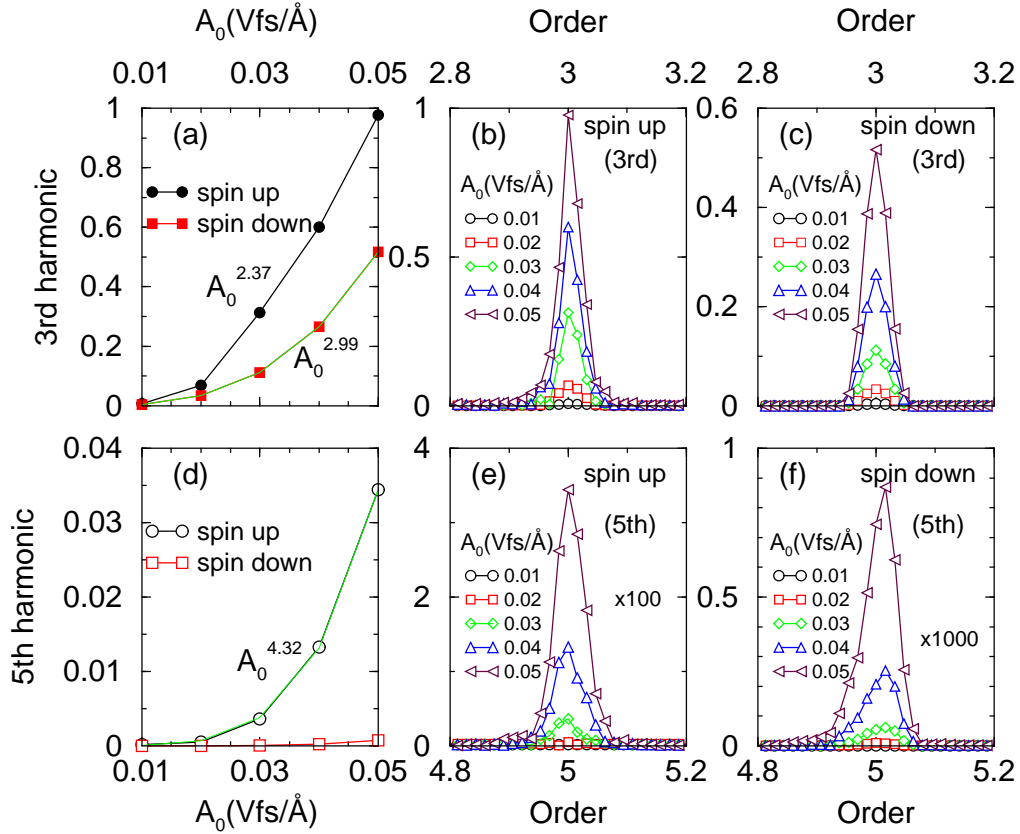}
\caption{Dependence of harmonic signals on the laser field amplitude
  $A_0$.  The photon energy is 0.4 eV, and pulse duration is 64 cycles
  of the laser period. Here we use a dense $k$ mesh of $29\times 29
  \times 45$. {(a) Third-order harmonics as a function of $A_0$
    for spin up (filled circles) and spin down (filled boxes), which
    are fitted to a power of $A_0^{2.37}$ and $A_0^{2.99}$.  (b) Third
    harmonic from the spin up channel. The circles, boxes, diamonds,
    upper triangles and left triangles denote $A_0=0.01$, 0.02, 0.03,
    0.04 and 0.05 $\rm Vfs/\AA$, respectively.  (c) Third harmonic
    from the spin down channel. All the legends have the same meaning
    as those in (b).  (d) Fifth-order harmonics as a function of $A_0$
    for spin up (filled circles) and spin down (filled boxes). The
    spin up signal can be fitted to a power of $A_0^{4.32}$.  (e) The
    fifth harmonic from the spin up channel. Here all the signals are
    multiplied by 100 for an easy view. All the legends have the same
    meaning as those in (b).  (f) Fifth harmonic from the spin down
    channel.  Here all the signals are multiplied by 1000. All the
    legends have the same meaning as those in (b).} }
\label{fig3}
\end{figure}

To show that our prediction is not specific to a particular laser
field strength, we choose several vector potential amplitudes. We
increase the amplitude $A_0$ from 0.01 to 0.05 $\rm Vfs/\AA$ and
include both the third and fifth harmonics. Figure \ref{fig3}(a)
displays the third harmonic signal as a function of $A_0$ for spin up
(filled circles) and spin down (filled boxes) channels.  The photon
energy is 0.4 eV and the duration is also 64 cycles of the laser
period as above. We see that both signals increase superlinearly. We
fit the signal to a power function and find that the signal scales
with $A_0$ as $A_0^{2.37}$.  According to the traditional nonlinear
optics \cite{shen}, the polarization ${\cal P}$ is proportional to the
response function $R(t)$ multiplied by the external field $E(t)$
\cite{mukamel}, and for the third harmonic ${\cal P}^{(3)}\propto
R^{(3)}(t) E(t)^3$. If $R^{(3)}(t)$ is significantly modulated by band
states, the scaling exponent differs from 3. This shows that real
excitations among band states contribute to the harmonic signal, which
is exactly what one expects from the density of states in the spin
majority channel (Fig. \ref{fig1}(a)). The spin minority channel is
different. We see that the signal scales as $A_0^{2.99}$ (filled boxes
in Fig. \ref{fig3}(a)), with the exponent close to 3. This again shows
that the excitation is virtual, as expected.

Different excitations seen above also leave a crucial hallmark on the
power spectrum as well. Figure \ref{fig3}(b) shows the third harmonic
signal as a function of $A_0$, from 0.01 to 0.05 $\rm Vfs/\AA$,
denoted respectively by the circles, boxes, diamonds, up triangles and
left triangles. We see that the peaks for the spin up channel are not
entirely symmetric with respect to the 3rd order. This is an
indication that harmonics are generated through real states
(Fig. \ref{fig1}).  In the spin down channel, the harmonic peak is
highly symmetric (see Fig. \ref{fig3}(c)). In other words, the
harmonic asymmetry is a hallmark for transitions that involve real
band states, which is crucial for electronic structure detection in
solids.

 We can further demonstrate this in the fifth harmonic. Figure
 \ref{fig3}(d) shows the 5th harmonic signal as a function of $A_0$
 for the majority spin (circles) and minority spin (boxes)
 channels. As we increase $A_0$, the signal scales as $A_0^{4.32}$,
 indicative of real transitions among band states.  To reveal further
 details in harmonics, we also plot the harmonic spectrum. Figure
 \ref{fig3}(e) shows that the harmonic peak is not symmetric with
 respect to 5.  All the harmonic signals are multiplied by 100 for an
 easy view.  In the spin down channel, we note that the nominal 5th
 peak does not appear at 5 (Fig. \ref{fig3}(f)), where harmonic
 signals are multiplied by 1000 for an easy view.  This is because the
 main excitation occurs above 2 eV, which is fully consistent with our
 observation in the photon energy dependence in Fig. \ref{fig2}.  In
 comparison with the majority signal, the minority signal is extremely
 weak even at 0.05 $\rm Vfs/\AA$. Quantitatively, we find the ratio
 between the majority and minority 5th order signals is 46, or about
 2\%.

\begin{figure}
\includegraphics[angle=0,width=0.8\columnwidth]{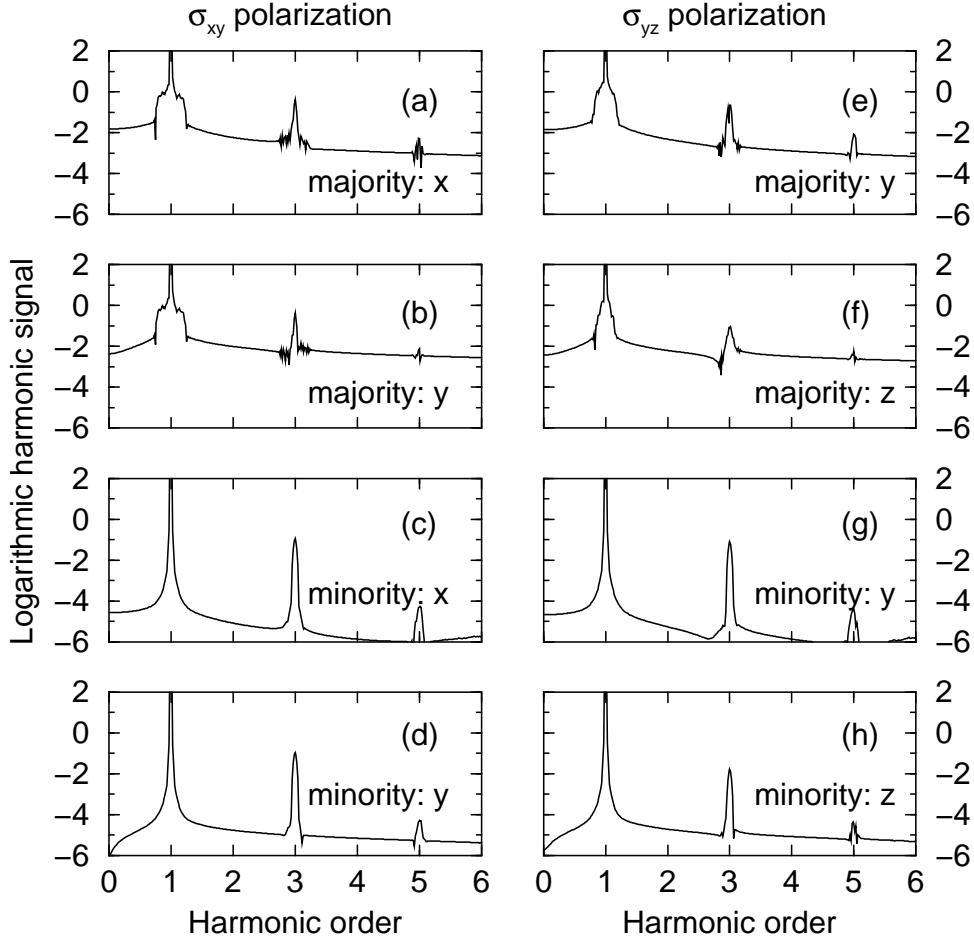}
\caption{{Influence of relaxation times and laser polarization on
    harmonics.  (a)-(d) use a different set of relaxation times,
    $T_1=400$ and $T_2=200$ fs with laser polarization within the $xy$
    plane.  (a) and (c) show the majority and minority harmonic
    signals along the $x$ direction respectively, which should be
    compared with those in Fig. \ref{fig1}(b) with $T_1=200$ and
    $T_2=100$ fs.  (b) and (d) show the majority and minority harmonic
    signals along the $y$ direction, respectively. (e)-(h) use the
    laser polarization in the $yz$ plane with $T_1=400$ and $T_2=200$
    fs.  (e) and (g) show the majority and minority harmonic signals
    along the $y$ direction, respectively.  (f) and (h) show the
    majority and minority harmonic signals along the $z$ direction,
    respectively.  For all the figures, the photon energy is 0.4 eV,
    the pulse duration is 64 cycles of the laser period and the laser
    field amplitude $A_0$ is 0.03 $\rm Vfs/\AA$.  } }
\label{fig4}
\end{figure}

{To establish our results on a solid ground, we carry out several
  additional tests.  We first check our harmonic signal convergence
  with the Brillouin zone sampling by increasing the number of $k$
  points from $18\times 18 \times 28$, $19\times 19 \times 30$, to
  $29\times 29 \times 45$, and we find the results are well converged.
  For all the results in this paper, we use a ${\bf k}$ mesh of
  $29\times 29 \times 45$, with 2760 points in the irreducible
  Brillouin zone, or 10830 in the full Brillouin zone.  Secondly, we
  examine the influence of relaxation times $T_1$ and $T_2$ on the
  harmonic signal.  $T_2$ is normally shorter than $T_1$, but their
  precise values are unknown, so to this end, we take $T_1=200$ fs and
  $T_2=100$ fs.  {\clr Extremely smaller $T_1$ and $T_2$ were used
    before in ZnO \cite{vampa2014}.}  Figures \ref{fig4}(a)-(d) show
  the results for the same laser parameters used in
  Fig. \ref{fig1}(a), but with $T_1=400$ fs and $T_2=200$ fs.  Their
  influence on our harmonic signal is very weak since our laser pulse
  duration is on the order of 200 fs.  When we increase $T_1$ and
  $T_2$ from 200 and 100 fs to 400 and 200 fs, the fifth order
  harmonic changes from $0.36\times 10^{-2}$ to $0.50\times 10^{-2}$
  for the majority channel.  The change is even smaller for the
  minority band, and it only changes from $0.55\times 10^{-4}$ to
  $0.52\times 10^{-4}$ for the minority channel in the same unit. This
  huge difference between the spin majority and minority channels
  remains, regardless of the $x$ (Figs.  \ref{fig4}(a) and \ref{fig4}
  (c)) or $y$ (Figs. \ref{fig4}(b) and (d) or Figs. \ref{fig4}(e) and
  \ref{fig4}(f)) or $z$ component (Figs. \ref{fig4}(f) and
  \ref{fig4}(h)).  Since \Cr is tetragonal, we wonder whether the
  laser polarization in the $yz$ plane ($\sigma_{yz}$), instead of the
  $xy$ plane ($\sigma_{xy}$), makes a difference. Figures
  \ref{fig4}(e)-(h) show our results with the same relaxation times as
  Figs.  \ref{fig4}(a)-(d).  We find that for the majority band, the
  $y$ component maximum of the 5th harmonic signal is similar, but the
  line shape is slightly different. Under $\sigma_{yz}$ excitation,
  the peak appears more symmetric (see Fig. \ref{fig4}(e)) and has no
  structure in it, while under $\sigma_{xy}$ excitation, the peak has
  more structures. This difference is due to the different band states
  involved during harmonic generation. This potentially provides
  another interesting direction for experimental investigation as a
  function of laser polarization angle, in a similar way as done in
  MgO \cite{you2017}. For the minority band, the difference is much
  smaller (see Figs. \ref{fig4}(g) and \ref{fig4}(f)). Importantly,
  the ratio of the majority 5th harmonic to that of the minority
  remains the same.}  This concludes that the fifth order harmonic
emission is mainly from the majority channel.  This paves the way to
future experimental detection of harmonics from a single spin channel.
We believe that the results found here are interesting and should
motivate future experimental and theoretical investigations.

\section{conclusion}

We have investigated harmonic generation from half-metallic \Cre. This
nearly 100\% spin polarized material shows peculiar features that are
not found in other materials. The majority spin channel has no energy
gap just like a metal, but in the minority channel has 1.4-eV energy
gap. Upon laser excitation, this leads to a significant difference in
harmonic generation in these two channels. This difference depends on
the harmonic order and photon energy $\hbar\omega$ used. The higher
the order is, the larger the effect is. At $\hbar\omega=0.4$ eV and
the fifth harmonic, we find that the ratio of the majority channel
signal to the minority signal exceeds 60. This makes the 5th harmonic
as if it is emitted {nearly} purely from a single spin
channel. Microscopically, in the minority channel, transitions are
mainly from the O-$2p$ states to Cr-$3d$ states, so the onset of
single spin channel harmonics starts at 2 eV. This explains why the
fifth order shows such a strong effect. In general, it is known that
harmonics are less sensitive to material properties. Here, we show
that when a material has a peculiar difference in spin channels,
harmonics can be very powerful.  Our prediction, if confirmed
experimentally, potentially represents a new direction for high
harmonic generation for solids
\cite{ghimire2011,luu2015,garg2016,nd,neufeld2019}.

\acknowledgments
This work was solely supported by the U.S. Department of Energy under
Contract No. DE-FG02-06ER46304. Part of the work was done on Indiana
State University's high performance Quantum and Obsidian clusters.
The research used resources of the National Energy Research Scientific
Computing Center, which is supported by the Office of Science of the
U.S. Department of Energy under Contract No. DE-AC02-05CH11231.

$^*$guo-ping.zhang@outlook.com. ~~  https://orcid.org/0000-0002-1792-2701

%\newpage
\clearpage

\begin{table}
\caption{Optimized structural parameters of \Cre.  Cr
  atoms are at $(2a)$ positions, (0,0,0) and
  $(\frac{1}{2},\frac{1}{2},\frac{1}{2})$, and O atoms at $(4f)$
  positions, ($u$,$u$,0), ($\bar{u}$,$\bar{u}$,0), ($\frac{1}{2}+u,
  \frac{1}2-u, \frac{1}2$), and
  $(\frac{1}{2}-u,\frac{1}{2}+u,\frac{1}{2})$.  Our theoretical values
  are listed under ``Theory'', with the experimental ones in
  parenthesis.  Our theoretical $u$ is 0.303, in good agreement with
  the experimental $u=0.301\pm 0.004$ \cite{cloud1962}.  The
  experiment lattice constants are $a=4.4218~\rm\AA$ and
  $c=2.9182~\rm\AA$ \cite{cloud1962}, $a=4.423~\rm \AA$ and
  $c=2.918~\rm \AA$ \cite{singh2009}, $a=4.421~\rm\AA$ and
  $c=2.916~\rm\AA$ \cite{pathak2009}, and $a=4.421~\rm\AA$ and
  $c=2.916~\rm\AA$ \cite{thamer1956}.  We adopt the experimental
  lattice constant from Cloud \et \cite{cloud1962}.  }
\begin{tabular}{c|c|c|c|c|c|c}
\hline
\hline
Atom & \multicolumn{2}{c|} {$x$} &  \multicolumn{2}{c|}{$y$}& \multicolumn{2}{c} {$z$} \\
\hline
      &~~Theory~~  &Experiment     &~~Theory~~&Experiment &~~Theory~~ &Experiment\\
Cr$_1$ &0 &(0)  & 0  &(0) &0 &(0)\\
Cr$_2$ & $\frac{1}2$ &($\frac{1}2$) & $\frac{1}{2}$ &($\frac{1}2$) & $\frac{1}{2}$
& ($\frac{1}2$) \\
O$_1$ &0.303 &(0.301) &0.303 &(0.301) &0& (0) \\
\hline\hline
\end{tabular}
\label{tab}
\end{table}

%Add the real time evolution of the majority and minority spins. 

\end{document}